\documentclass[aps,showpacs,showkeys,preprint]{revtex4}
\usepackage{graphicx}
\usepackage{slashed}

\begin{document}{\normalsize}

\title{Subthreshold pair production in short laser pulses} 

\author{T. Nousch}
\author{D. Seipt}
\author{B. K\"ampfer}

\address{Helmholtz-Zentrum Dresden-Rossendorf,
POB 510119, 01314 Dresden, Germany\\
TU Dresden, Institut f\"ur Theoretische Physik, 01062 Dresden, Germany}

\author{A.~I.~Titov}
\address{Bogoliubov Laboratory of Theoretical Physics, JINR, Dubna 141980, Russia\\
Institute of Laser Engineering, Yamada-oka, Suita, Osaka 565-0871, Japan}

\begin{abstract}
The $e^+ e^-$ pair production by a probe photon traversing a linearly polarized
laser pulse is treated as generalized nonlinear Breit-Wheeler process. 
For short laser pulses with very few oscillations of the electromagnetic field we find
below the perturbative weak-field threshold $\sqrt{s} = 2m$ 
a similar enhancement of the pair production rate as for circular polarization.
The strong subthreshold enhancement is traced back to the finite bandwidth of the
laser pulse. A folding model is developed which accounts for the interplay of the
frequency spectrum and the intensity distribution in the course of the pulse.
\end{abstract}


\maketitle

\section{Introduction}

The Breit-Wheeler process [1],
$\gamma' + \gamma \to e^+ + e^-$, describes pair production as a $t$ channel
process, which is sometimes termed "conversion of light into matter". Considering it as
2-to-2 process in a particle physics language, the Breit-Wheeler pair creation is a threshold
process, i.e.\ the available energy, expressed by the Mandelstam variable 
$s = (k' +k)^2 =2 \omega' \omega (1 - \cos \Theta_{\vec k' \vec k})$, 
must fulfill $s > 4 m^2$, where $m$ is the electron mass. Supposed, $\gamma$
describes optical (laser) photons with energy $\omega \sim 1$ eV, 
the energy of the counter propagating
photon $\gamma'$ must obey $\omega' > 250$ GeV. While such high energy photons might be generated
by various processes at high energy accelerators/colliders, e.g.\ in the time reversed 
Breit-Wheeler process (pair annihilation) or Compton backscattering of a laser beam off a very
energetic electron beam, or in astrophysical environments, the availability of laboratory
based 250 GeV photon sources is quite scarce. Nevertheless, the Breit-Wheeler process has
been identified experimentally. The SLAC experiment E-144 \cite{E-144} was a special set-up of
one of the above mentioned options.

In a sufficiently strong laser field, however, multi-photon processes are enabled
\cite{Reiss,NR-64,Ritus-79,Serbo}, 
schematically $\gamma' + n \gamma \to e^+ + e^-$, often termed nonlinear Breit-Wheeler
process. Indeed, the interpretation \cite{Ilderton-2010} of the SLAC experiment E-144 has
proved this possibility. In fact, at least $n = 4$ laser photons were needed to produce a pair.
For an all-optical set-up with $\omega' \sim \omega$, $n > 10^{11}$ laser photons 
would be required when replacing in the above threshold formula $k \to nk$. 
Strong laser fields are necessary for such nonlinear effects. 
The electron and positron in a strong laser field acquire an effective
mass, $m_* = m \sqrt{1 + a_0^2/2}$, where $a_0$ 
is the dimensionless  laser-strength parameter 
(cf.~\cite{Heinzl-PLB,Hebenstreit,etc} 
for a recent analysis of the mass dressing in strong pulsed laser fields). 
That is, the above threshold estimate would read $s > 4m_*^2$
in disfavor of achieving the threshold for strong optical laser fields. The up-shift of the
threshold, however, is compensated to some extent by the higher harmonics with
thresholds $s_n = 4 m_*^2 / n$. That means, in principle, the higher harmonics
(labelled by $n$)
seem to shift the threshold towards arbitrarily small values of $s$. 
For a given value of $s$, the minimum number of laser photons $n_0$ is the smallest
integer such that $s_{n_0} \le s$.   
In the perturbative regime, where $a_0 \ll 1$, the harmonics are suppressed by 
factors of $a_0^{2n}$, and a noticeable pair production therefore
extents not too far below
the 2-to-2 threshold of $4 m^2$. 
In the deep subthreshold region $a_0 \gg 1$ and $\kappa \ll 1$, 
a large number of laser photons participate in the formation of pair
with cross section $\sigma \propto e^{-8/(3 \kappa)}$, where 
$\kappa = s a_0/(2 m^2)$ is the Ritus parameter for pair production   
\cite{Ritus-79}.
It has a similar non-analytic
dependence on the field strength parameter encoded in $\kappa$ as the Schwinger effect 
which depends on the electric field amplitude (cf.~\cite{Ringwald}.)
At fixed $a_0$, there is a steep decline of the pair creation probability for
$s < 2 m^2 / a_0$, i.e.\ large values of $a_0$ shift the region of noticeable
pair production to small values of $s$.

The cross section for pair production in the nonlinear Breit-Wheeler process
by an unpolarized probe photon $\gamma'$
in a plane wave with linear polarization reads in such a case \cite{Ritus-79}
$\sigma (s) = \sum_{n \ge n_0}^\infty \sigma_n (s)$ with
\begin{eqnarray}
\sigma_n (s) &=& \frac{4 \alpha^2}{s a_0^2}
\int_0^{2 \pi} d\varphi \,  
\int_1^{u_n^*} \frac{du}{u \sqrt{u (u-1)}}
\left\{ A_0^2 + a_0^2 (2 u -1 ) \left( A_1^2 - A_0 A_2 \right) \right\},  \label{eq.1}
\end{eqnarray}       
where 
$A_j \equiv A_j(n, a, b) =
\frac{1}{2\pi} \int_{- \pi}^{\pi} dt \, \cos^j t \, 
\exp\{ i (a \sin t - b \sin2 t - n t) \}$
are generalized Bessel functions with arguments $n$,
$a =  \frac{8 m^2}{s} a_0 \sqrt{1 + a_0^2/2} \, \sqrt{u (u_n^* -u)} \, \cos \varphi$, 
and $b = a_0^2 u m/s$; $u_n^*$ is defined by
$u_n^* = n \frac{s}{4 m_*^2}$.
Note that the relevant energy variable is still $s = (k' + k)^2$.
In Fig.~1, a survey of the cross section is exhibited
as a contour plot over the $\sqrt{s}$ vs.\ $a_0$ plane. 
The apparent structures 
visible in the region $0.8 < \sqrt{s} / {\rm MeV} < 1$ for $a_0 < 1$ 
are related to the turn over from the first to the second harmonic.
The features propagate to increasing values of $\sqrt{s}$ for $a_0 > 1$. 
The imprints of the onset of the second and third harmonics are still visible.
In the weak-field region, $a_0 < 1$, 
the steep decline of the cross section as a function of $\sqrt{s}$ is evident below  
$\sqrt{s} = 2 m$ .

\begin{figure}[h]
\includegraphics[width=0.69\textwidth]{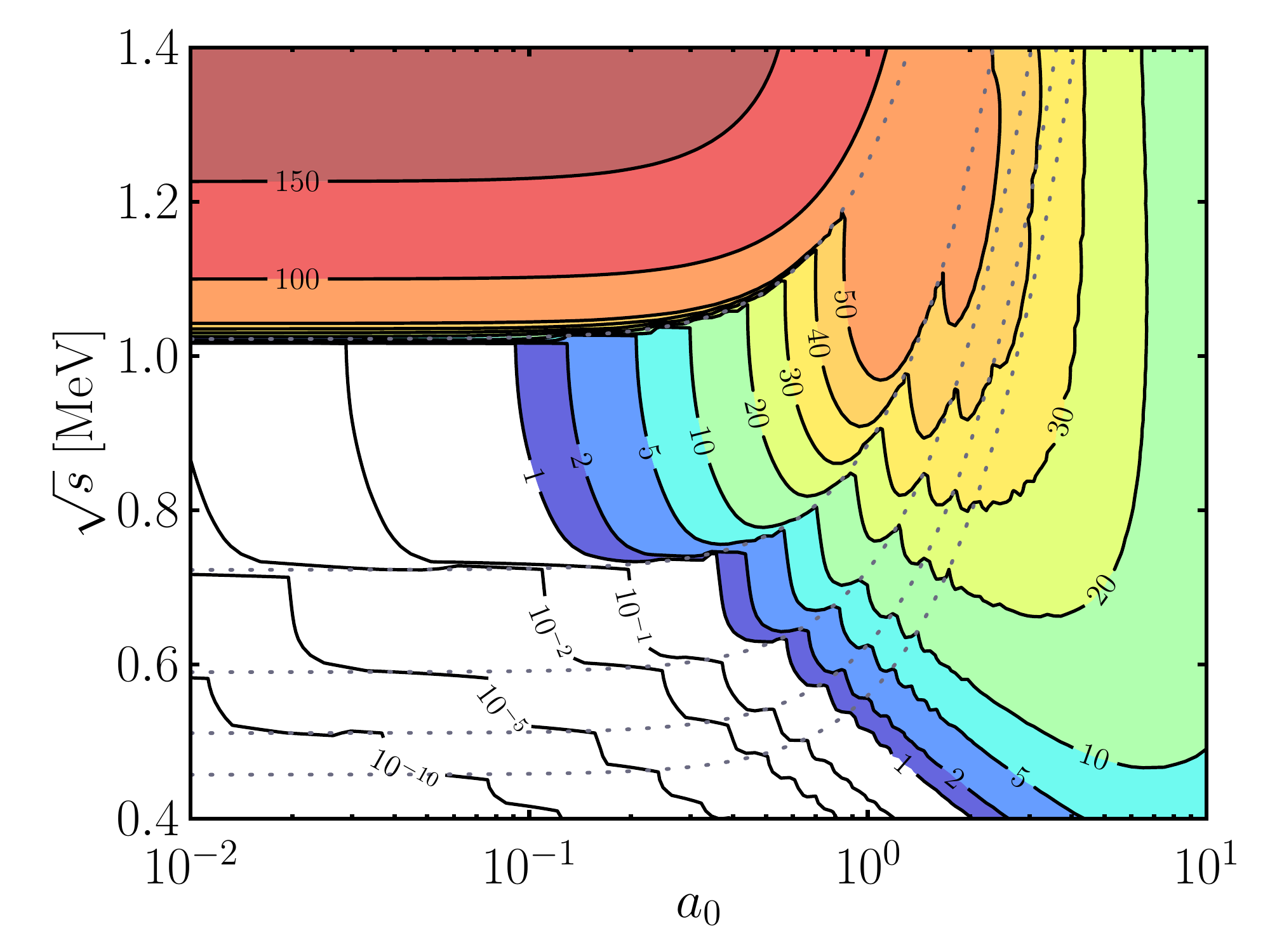}
\caption{Contour plot of the cross section (\ref{eq.1})
over the $\sqrt{s}$ vs.\ $a_0$ plane. The numbers label curves of constant
cross sections in mb. Dashed curves depict the thresholds $s_n$
for $n = 1 \cdots 5$ (from top to bottom). 
\label{fig1}}
\end{figure}

In the case of circular polarization, Eq.~(\ref{eq.1}) becomes
modified \cite{Ritus-79,Serbo}. The overall pattern remains, but with less
pronounced structures due to azimuthal symmetry.
In the strong-field region, $a_0 \gg 1$, the subthreshold production with a large 
number of photons is suppressed 
due to the large angular momentum of the pair \cite{Ritus-79}.

The hitherto discussed examples are for infinitely extended plane waves.
In reality, strong laser fields are generated presently by the chirped
pulse amplification technique,
i.e. pulse compression, and the asymptotic final state refers to a free electron-positron pair,
where $m_*$ does not matter.
Therefore, a modified threshold behavior is to be expected. 
Moreover, a pulse of finite duration is not longer monochromatic, instead the power
spectrum has a support of finite width. Also this effect will have imprints on the
pair production, as does the variation of the intensity in the course of the pulse.
Considering the (nonlinear) Breit-Wheeler process as cross channel of the
(nonlinear) Compton scattering one can expect similar strong effects 
of the temporal pulse shape, as has been found for the latter one
\cite{Boca-2009,Mackenroth-2011,NF-96,Seipt-2011}. 
In fact, the differential $e^+ e^-$ spectra in the (general) Breit-Wheeler process
depend sensitively 
on the laser pulse shape \cite{Heinzl-PLB}. Even for weak laser intensities $a_0 < 1$, 
a significant enhancement of the total pair production rate just below the threshold
$\sqrt{s} = 2 m$ has been found recently \cite{Titov} for short circularly polarized pulses. 

Given this motivation we study here the pair production off a probe photon in a 
linearly polarized laser pulse. 
In section 2, we recapitulate the basic formulas within the Furry picture,
where the interaction with the probe photon $\gamma'$ is treated perturbatively, while
the interaction of the electron and positron with the laser pulse is accounted for
by Volkov states. The numerical results are discussed in section 3
for the total pair production probability.
A folding model is introduced in section 4 to access to the effect
of enhanced subthreshold pair production, thus identifying the relevant physical
mechanisms. The summary is given in section 5.    

\section{Pair production in pulsed laser fields}

The pair creation process is a crossing channel of the Compton scattering. Accordingly,
we evaluate the cross section of the process
as decay of a probe photon $\gamma'$ 
with four-momentum $k'$ and polarization four-vector $\epsilon'$ 
into a pair $e^+(\gamma) + e^- (\gamma)$, where
$e^\pm (\gamma)$ are laser dressed Volkov states  (cf.~\cite{LandauLifshitz4}) 
which encode the interaction with the laser field.
For weak laser fields, $a_0 \ll 1$,
this process may be resolved perturbatively into a series of diagrams, where,
in addition to the probe photon $\gamma'$, $n$ laser photons $\gamma$ interact with the outgoing
$e^\pm$. The diagram with $n = 1$ at $a_0 \ll 1$ reproduces the Breit-Wheeler process.

The linearly polarized laser pulse is described here by the real four-vector potential
\begin{equation}
A^\mu (\phi) = \frac{m a_0}{e} \, \epsilon^\mu g(\phi)\, \cos \phi
\label{eq.pot}
\end{equation}
with transverse polarization four-vector $\epsilon^\mu$
obeying $k \cdot \epsilon = 0$; $e$ denotes the elementary charge.
The envelope function $g(\phi)$,
with $\phi = k \cdot x$ as invariant phase, encodes the temporal shape of the pulse
with wave four-vector $k$ and normalization $g(0) = 1$.
We chose in this paper, following \cite{Heinzl-PLB}, 
\begin{equation} \label{pulse}
g(\phi) = \cos^2 \left( \frac{\phi}{2 N}\right),
\end{equation}
for $\vert \phi \vert \le \pi N$ and $g(\phi) = 0$ for $\vert \phi \vert > \pi N$,
where $N$ is number of cycles in the pulse.
In line with the notation in \cite{Titov} we denote
the considered process as finite pulse approximation (FPA), in contrast
to the infinite pulse approximation (IPA) where $g \to 1$. Both, IPA and FPA,
use plane waves, i.e.\ $\epsilon \cdot k = 0$.

The $S$ matrix for the generalized nonlinear Breit-Wheeler process is   
\begin{eqnarray}
S & = & -ie\int d^{4}x\bar{\Psi}_{p'}(x)\slashed\epsilon^{\prime}
\frac{e^{-ik^{\prime}\cdot x}}{\sqrt{2\omega^{\prime}}}\Psi_{-p}(x),\label{eq:S_matrix}
\end{eqnarray}
where $\bar \Psi_{p'}$ and $\Psi_{-p}$ denote the Volkov wave functions 
(to be taken with the potential (\ref{eq.pot}))
of the outgoing electron and positron with four-momenta $p'$ and $p$,
respectively.
For instance, the positron wave function is given by
\begin{eqnarray}
\Psi_{-p}(x) & = & \left(1-\frac{e}{2p\cdot k}\slashed k\slashed A(\phi)\right)
e^{ip\cdot x-if_{-p}(\phi)}\frac{v_{p}}{\sqrt{2p_{0}}},\\
f_{-p}(\phi) & = & \frac{1}{2p\cdot k}\intop_{0}^{\phi}d\phi'
[2 e p \cdot A(\phi') + e^{2}A^{2}(\phi')]
\end{eqnarray}
with outgoing free-field spinor $v_{p}$.
Feynman's slash notation is used, e.g. $\slashed A = \gamma_\mu A^\mu = \gamma \cdot A$,
where $\gamma$ stands for the Dirac matrices.
We employ the light cone coordinates $x_{\pm}=x^{0}\pm x^{3}$ and
$\mathbf{x}_{\perp}=(x^{1},x^{2})$ and orient the coordinate system to have
$k \sim (\omega, 0, 0, -\omega)$, i.e.\ $k_- = 2 \omega$, $k_{+, \perp} = 0$ and $\phi = k_-x_+/2$. 
Therefore, the integration measure is $d^4 x = dx_- d^2 x_\perp  d \phi /k_-$.
After integrating over the components $x_{-}$ and $\mathbf{x}_{\perp}$
the matrix element reads 
\begin{eqnarray} \label{eq.8}
S & = & \frac{-ie}{\sqrt{2k_{0}^{\prime}2p_{0}2p_{0}^{\prime}}}
(2\pi)^{4} \int \frac{d \ell}{2\pi} \delta^{(4)} (p + p' - k' - \ell k) \,
{\cal M}(\ell),\\
{\cal M}(\ell) & = & \bar{u}_{p^{\prime}}\slashed\epsilon^{\prime}v_{p}\, B_0(\ell) 
+ m a_0
\bar{u}_{p^{\prime}}\Big( 
\frac{\slashed\epsilon \slashed k\slashed\epsilon^{\prime}}{2 k\cdot p^\prime}
- \frac{\epsilon^{\prime} \slashed k\slashed\epsilon}{2 k \cdot p} \Big)
v_{p} \,B_1(\ell) 
- \frac{m^2 a_0^2}{2 k\cdot p k\cdot p^\prime}\bar{u}_{p^{\prime}}\slashed kv_{p} \, B_2(\ell)
\end{eqnarray}
with the three phase integrals 
\begin{eqnarray} \label{eq.10}
B_m(\ell) &=& \int d\phi \, g^{m}(\phi) \, \cos^m (\phi) \,
\exp\left\{ i \left(\ell \phi - f_{p'}(\phi)+f_{-p}(\phi) \right) \right\} . 
\end{eqnarray}
The integral $B_0(\ell)$ does not contain an envelope function in the preexponential
and needs a regularization prescription, e.g.~as proposed in \cite{Boca-2009}.
The energy-momentum conservation in (\ref{eq.8}) fixes the value
$\ell \equiv \ell_0 = (p_{-}+p'_{-}-k'_{-})/k_{-}$. 
The pair emission probability reads with $\alpha = e^2 /(4\pi)$
\begin{eqnarray}
dW & = & \frac{\alpha}{8\pi^{2}(k\cdot k')(k\cdot p')}|{\cal M}|^{2}
p_{\perp}dp_{\perp}dyd\varphi.\label{eq:dW}
\end{eqnarray}
Here, we employ the rapidity 
$y=\frac{1}{2}\ln(p^{0}+p^{3})/(p^{0}-
p^{3}) = \frac12 \ln p_+ / p_- = \ln p_+ / m_\perp$
and transverse momentum $p_{\perp}=\sqrt{p_{1}^{2}+p_{2}^{2}}$ to
parameterize the final state phase space of the positron;
$m_\perp = \sqrt{m^2 + p_\perp^2}$ is the transverse mass. 
Due to the
finite laser pulse in FPA, the variables $p_{\perp}$ and $y$ are independent.
That means, the three variables $p_\perp$, $y$ and $\ell$ are related by
$m_\perp (y, \ell) = \frac{2 \ell \omega}{\ell e^y + e^{-y}}$.
In IPA (cf.~\cite{Ritus-79}), i.e.\ for $g \to 1$, due to the periodicity of the
laser pulse, the phase integrals $B_m$ collapse to a $\delta$ comb and
$\int \frac{d \ell}{2\pi}$ in (\ref{eq.8}) changes into a sum
over harmonics $n$ with $\ell \to \ell_n = n - \frac 12 m^2 a_0^2 /(k \cdot p)
- \frac12 m^2 a_0^2 /(k \cdot p')$. Inserting $\ell_n$ into $m_\perp (y, \ell_n)$ 
and solving for $p_\perp$ one recovers the usual harmonics in IPA.

\begin{figure}[ht]
\includegraphics[width=0.69\textwidth]{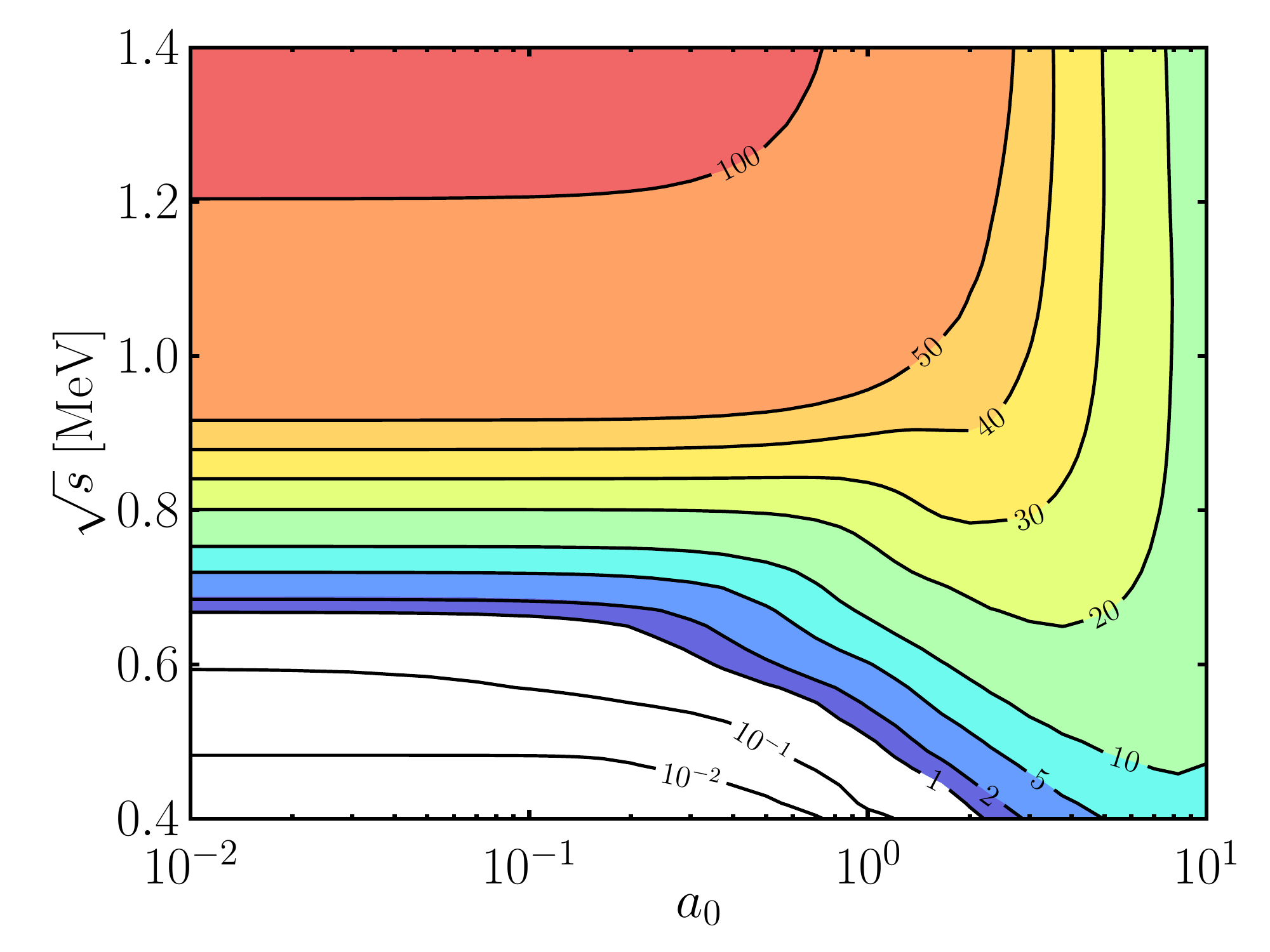}
\caption{
Contour plot of the cross section for linear polarization
over the $\sqrt{s}$ vs.\ $a_0$ plane. The numbers label curves of constant
cross sections in mb. FPA with $N = 1$.
\label{fig3}}
\end{figure}

\section{Total cross section \label{sQED}}

The numerical evaluation of the probability Eq.~(\ref{eq:dW}) leads to the cross section
$\sigma = W /(J \rho_\gamma)$, $J=2$ and 
$\rho_\gamma = m^2 a_0^2/(2\pi \alpha) \Delta \phi =
\omega^{-2} \int_{- \infty}^{\infty} d\phi \, T^{00}$,
where $T^{00}$ is the energy density of the laser background field, thus defining
an equivalent effective laser pulse duration $\Delta \phi$. 
For not too short pulses, where the approximations of \cite{Titov} are suitable,
one finds $\Delta \phi = \int_{- \infty}^{\infty}  d \phi \, g^2(\phi)$.
These definitions ensure
that the cross section is always based on the same energy in the laser pulse
irrespectively of its duration.      
The cross section is exhibited in Fig.~\ref{fig3} for $N = 1$, i.e., only
one cycle in the pulse, as contour plot over the $\sqrt{s}$ vs.\ $a_0$ plane.

The white region, where $\sigma < 1$ mb,
is completely changed in FPA (Fig.~\ref{fig3})
in comparison with the IPA case (Fig.~\ref{fig1}).
For $a_0 < 1$, significant pair production occurs at smaller
values of $\sqrt{s}$. The reason is explained in \cite{Titov}.
In IPA, the $n$th harmonic contributes only for $s \ge s_n$; in FPA, it
has a finite contribution also at $s < s_n$. This effect is most dramatic for
small values of $a_0$, where only the lowest harmonics contribute.

\begin{figure}
\includegraphics[width=0.49\textwidth]{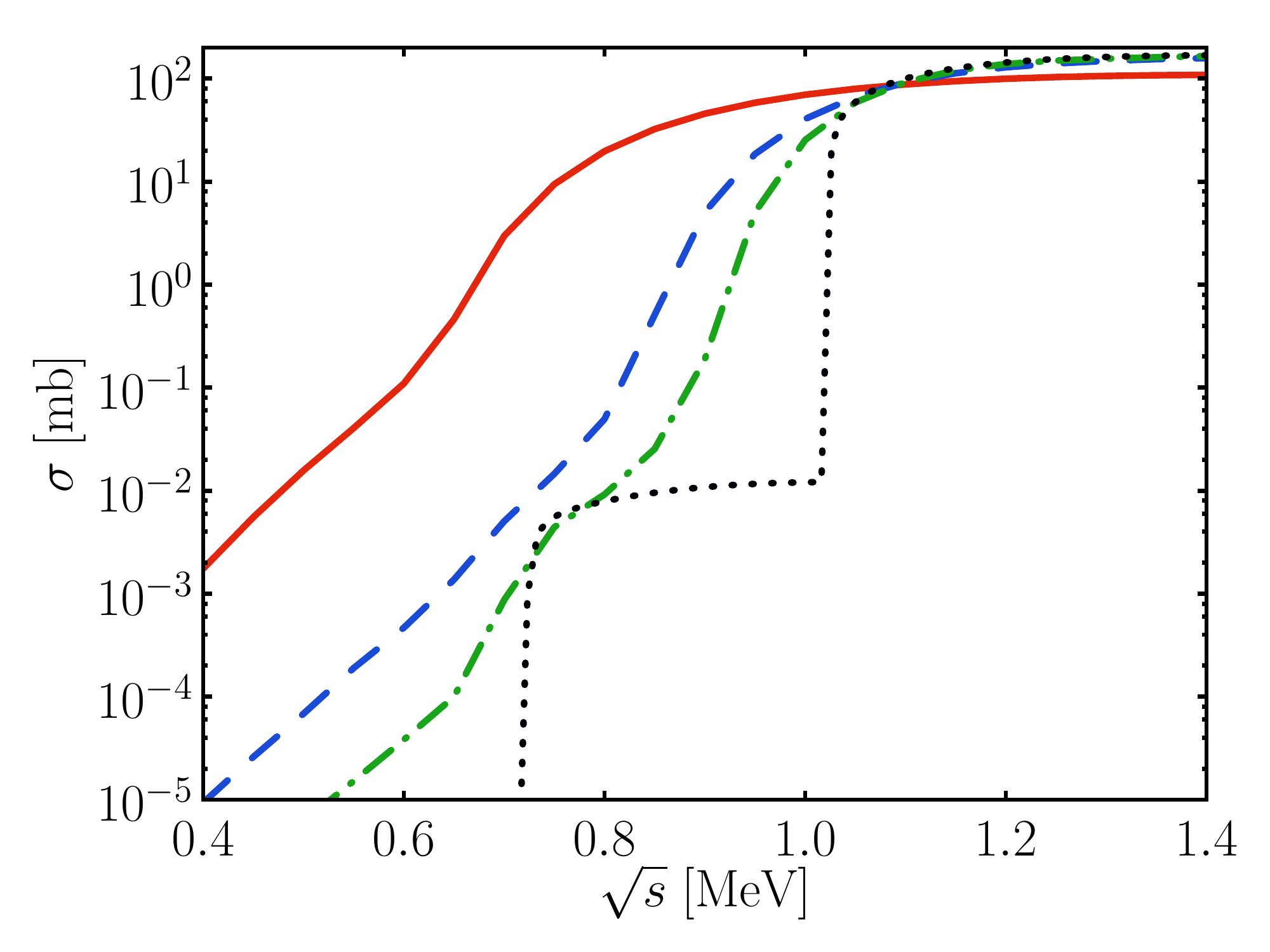}
\includegraphics[width=0.49\textwidth]{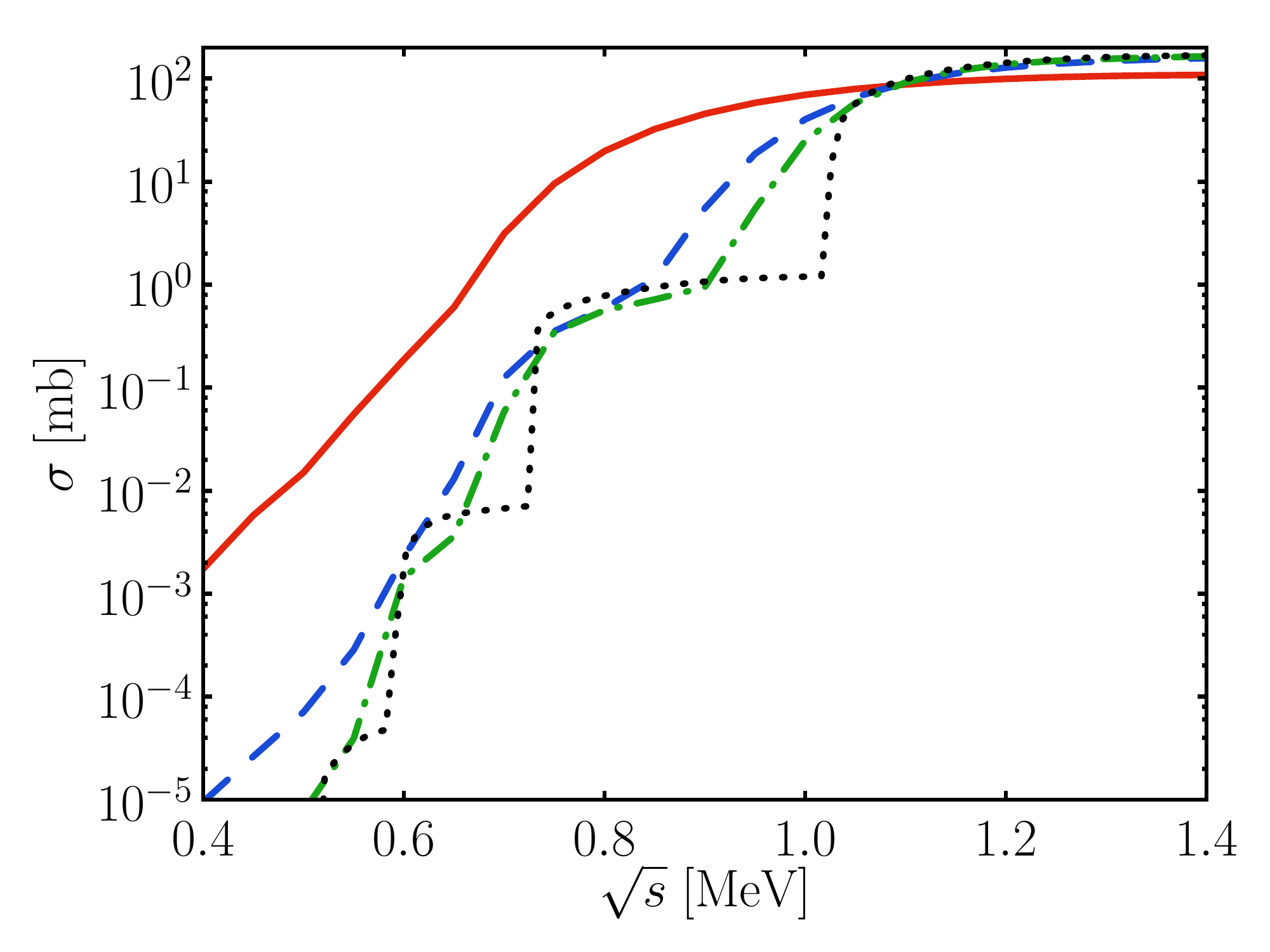}
\caption{Cross sections as a function of $\sqrt{s}$ for $a_0 = 0.01$ (left) and
0.1 (right).
The dotted curves are for IPA, while the 
other curves are for FPA with $N = 1$ (solid red), 3 (dashed blue) and 5 (dot-dashed green). 
\label{fig4}}
\end{figure}

To show the influence of the number of cycles in the pulse, we exhibit in
Fig.~\ref{fig4} the cross sections $\sigma (\sqrt{s})$
for $N = 1, 3$ and 5 and compare with the
IPA case for two values of $a_0$. The enhancement of the subthreshold
production can be many orders of magnitude above the IPA value at small
values of $a_0$, in particular for very short pulses. 
The enhancement is stronger than that found in \cite{Titov} for circular polarization,
e.g.\ for $a_0 = 0.01$, our $N = 5$ curve exceeds the second IPA harmonic by a factor
of 20 at $\sqrt{s} = 0.9$ MeV (see left panel of Fig.~\ref{fig4}), 
while in \cite{Titov} the $N = 5$ curve is even slightly below the IPA curve
at the same value of $\sqrt{s}$.    
(This can be attributed mainly to the different envelope function in \cite{Titov}.)
For $a_0 \ge 1$,
the pulse length does not matter for the chosen pulse shape: The differences
of IPA and FPA are within a factor of two.

\section{Folding the power spectrum with harmonics}

The above discussed enhancement of subthreshold pair production for $a_0 \ll 1$
is explained
for a circularly polarized laser beam in \cite{Titov} by exploiting properties
of the special functions entering the expressions for the $S$ matrix element
or cross section, respectively. Instead of transferring the formalism of \cite{Titov}
to linear polarization, we apply here a simple folding model to explain 
in an alternative and more intuitive manner the effect. 
The key is the finite bandwidth of a pulse:
The power spectrum contains frequencies lower and larger than the central value
$\omega$. This is accounted for in our model by 
calculating the weighted average of the
pair production cross section of all energy components which are present in the
laser pulse. Thus, a laser pulse with larger bandwidth will lead to an increased 
subthreshold production due to the high-frequency content.
Denoting by  $G(\ell)$ the Fourier transform of the pulse envelope,
$G(\ell) = \int d\phi \, g(\phi) \, e^{i \ell \phi}$,
the weighted cross section for the $n$th harmonic can be defined as
\begin{equation}
\langle \sigma_n \rangle (s) = R_n \frac{\int_0^\infty d\ell \, G(\ell - 1)^{2n} 
\sigma_n^{(0)}(\ell s)}{\int_0^\infty ds \, G(\ell-1)^{2n}} , \label{eq.convolution} 
\end{equation}
with the IPA cross sections $\sigma_n^{(0)}$ 
in the weak-field (perturbative) limit, i.e., the leading term in an expansion
in powers of $a_0$. For instance, $\sigma_1^{(0)}$ is the Breit-Wheeler cross section 
(cf.~\cite{LandauLifshitz4})
\begin{equation} \label{BW1}
\sigma_1^{(0)}(s) = \frac{2 \pi \alpha^2}{s} 
\Big[ 
    (3-\beta_1^4) \ln \frac{1+\beta_1}{1-\beta_1} - 2 \beta_1 (2-\beta_1^2)
\Big] \label{eq.BW} ,
\end{equation}
and the second 
harmonic reads for linear polarization \cite{Ritus-79}
\begin{eqnarray} \label{BW2}
\sigma_2^{(0)} (s)&=& \frac{a_0^2}{4} \frac{2 \pi \alpha^2}{s} 
\left[
    \left(
    6 + \frac{3}{u_2} - \frac{20}{u_2^2} + \frac{15}{u_2^3} 
    \right) \beta_2
 + \frac{15}{2u_2^2} \beta_2^4 \ln \frac{1+ \beta_2}{1-\beta_2}
\right]\,, 
\end{eqnarray}
where $u_n = ns /(4m^2)$ and 
$\beta_n  = 
\sqrt{1 - u_n^{-1}}$.
In (\ref{eq.convolution}), 
$R_n = \int g^{2n} (\phi) d \phi / \int g^2 (\phi) d\phi$ is an average
over the intensity of the pulse relevant for the $n$th harmonic. It depends
on the pulse shape but is independent of the pulse length. (For instance, for the
pulse (\ref{pulse}), we find $R_2 = 0.72916$ and $R_3 = 0.6016$,
while for a Gaussian pulse shape one has $R_n = 1/\sqrt{n}$; the $\cosh$ pulse
of \cite{Titov} is characterized by $R_2 = 2/3$ and $R_3 = 8/15$; $R_1 = 1$
by definition.)
That means, the relative importance of different harmonics depends on the shape
of the pulse. 
If the frequency distribution does not matter, such as for long pulses with
$N \gg 1$, the effect of the intensity distribution for $a_0 \ll 1$
causes $\langle \sigma_n \rangle (s) \to R_n \sigma_n^{(0)}$ in (\ref{eq.convolution}).
The effect of the intensity distribution in the course of a pulse
leads to an overall reduced subthreshold production, while the frequency spectrum can lead
to an increased subthreshold production in the vicinity of the corresponding
thresholds $s_n$. This explains why FPA is below IPA for 
certain values of $\sqrt{s}$ in Fig.~\ref{fig4}
(e.g., at $\sqrt{s} \sim 0.8$ MeV in the right panel for $N = 3$ and 5). 

\begin{figure}[!t]
\begin{center}
\includegraphics[width=0.49\textwidth]{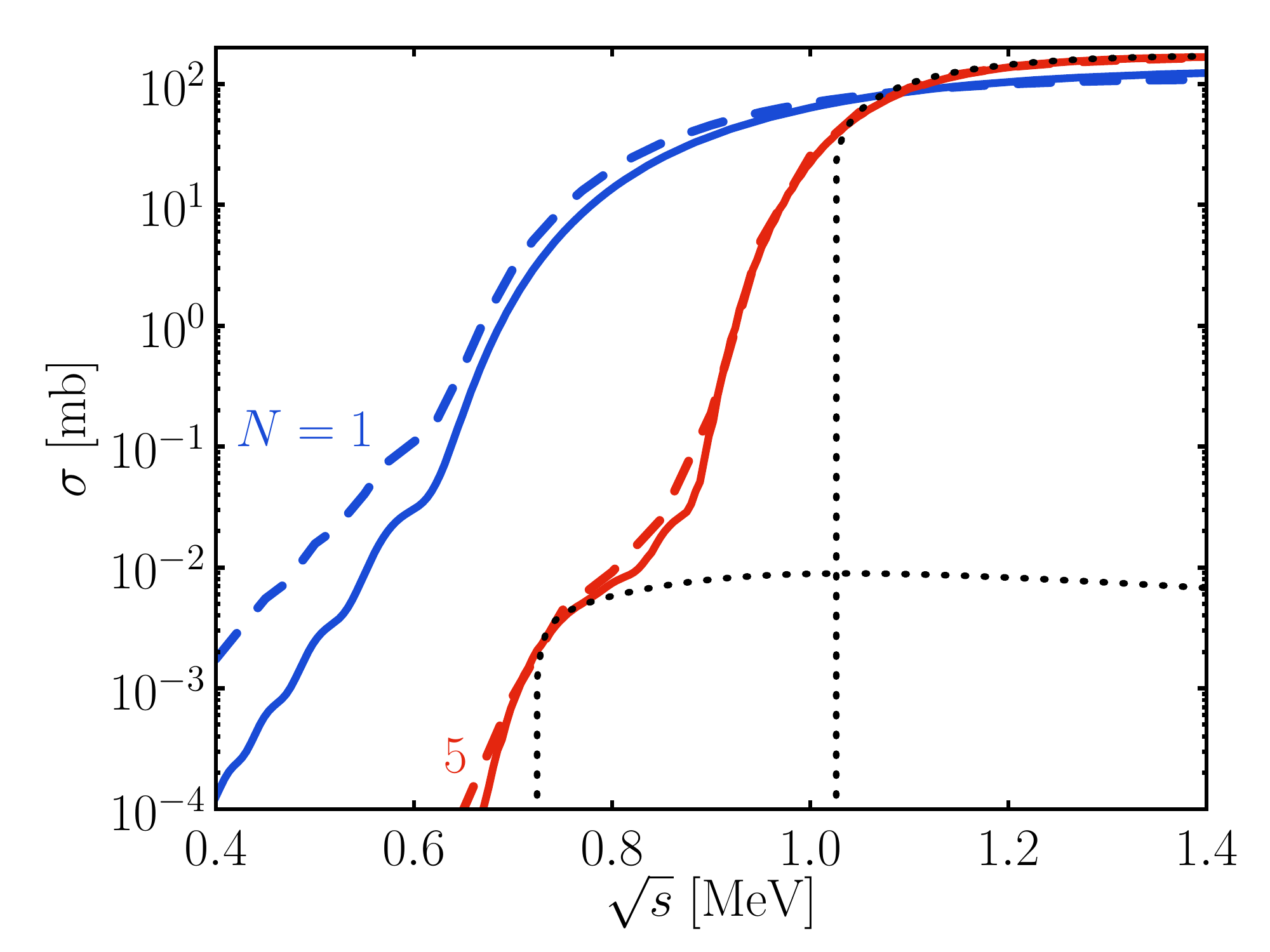} 
\includegraphics[width=0.49\textwidth]{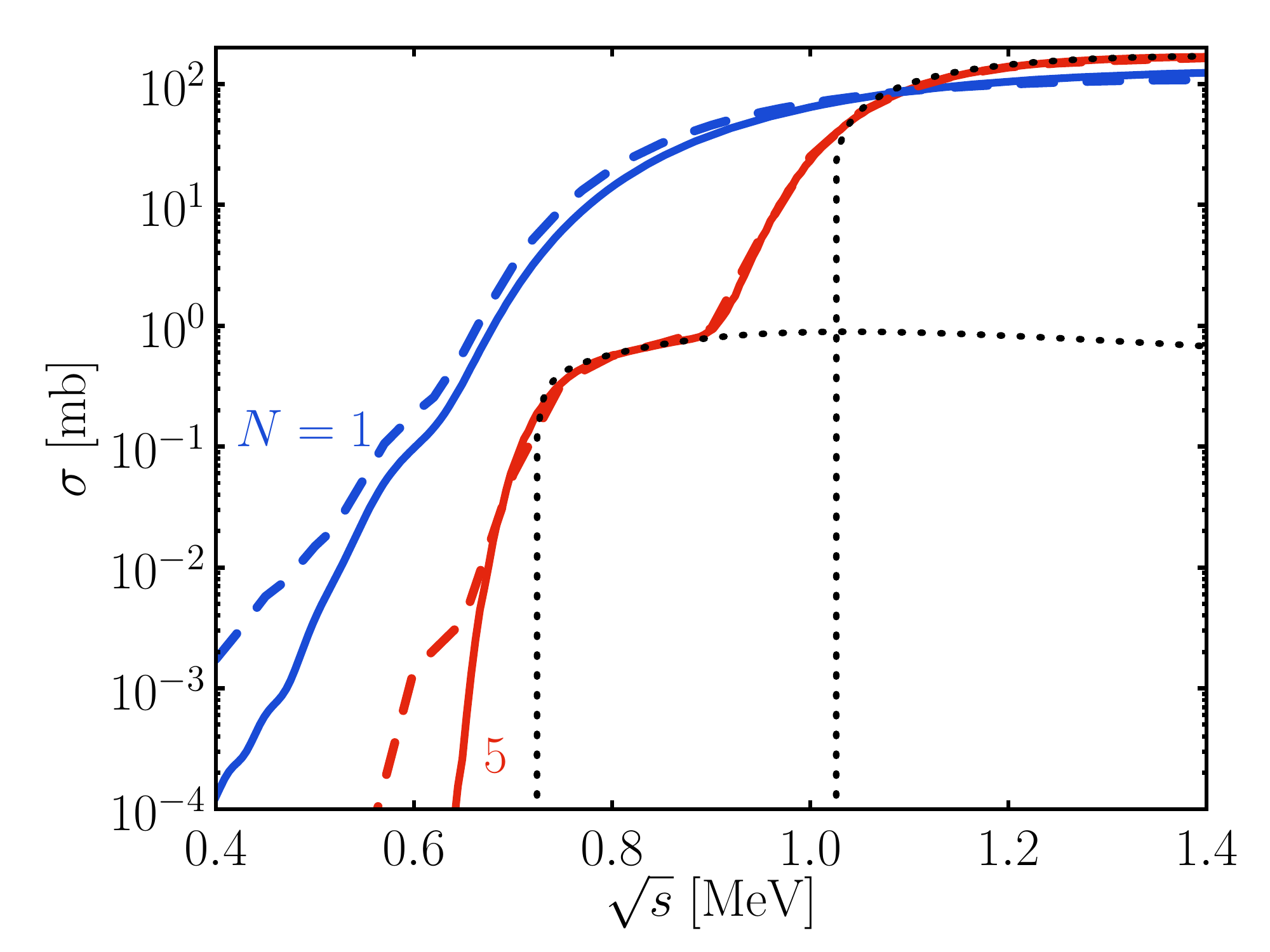}
\end{center}
\caption{Comparison of the folding model Eq.~(\ref{eq.convolution})
(solid curves for $\sigma = \langle \sigma_1 \rangle + \langle \sigma_2 \rangle$,
i.e. only the first two harmonics are taken into account) 
and the full numerical calculations from section \ref{sQED} (dashed curves).
The black dotted curves depict the modified nonlinear Breit-Wheeler cross sections 
$\sigma_n^{(0)} R_n$ for $n = 1$ and 2 (upper and lower curves).
Left panel: $a_0=0.01$, right panel: $a_0=0.1$.}
\label{fig.model}
\end{figure}

The comparison of that folding model with the above full numerical calculations 
is exhibited 
in Fig.~\ref{fig.model}. Despite of the simplicity of the folding model, 
the quantitative agreement is surprisingly good.  
We assign the deviations at small values of $\sqrt s$ to the omission 
of higher harmonics.
     
\begin{figure}[!t]
\begin{center}
\includegraphics[width=0.49\textwidth]{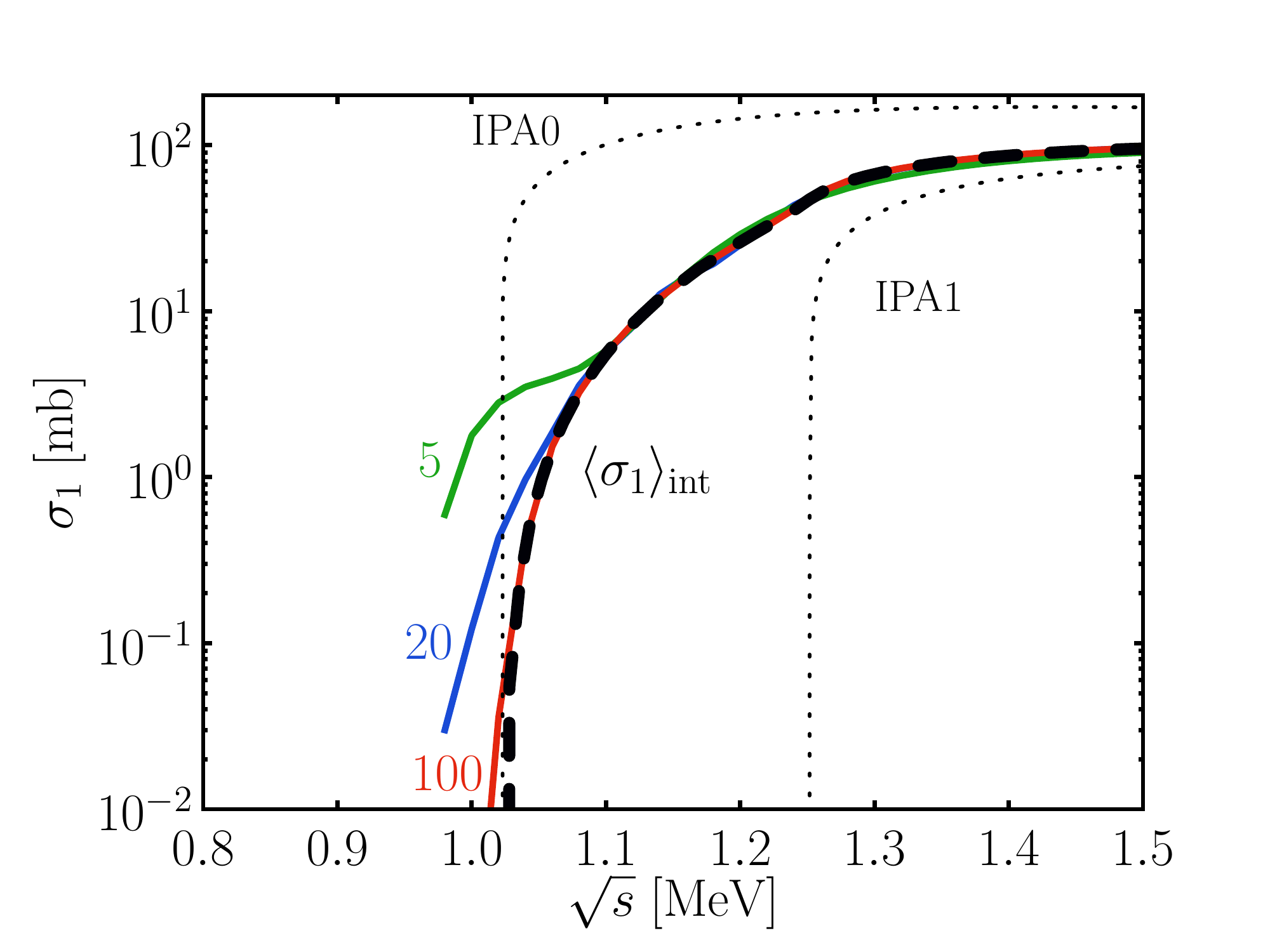}
\end{center}
\caption{$\sigma_1^{FPA}$ as a function of $\sqrt{s}$ for various values of $N$
(solid curves). The dashed curve depicts $\langle \sigma_1 \rangle_{\rm int}$
for $a_0 = 1$. The dotted curves are the IPA results for the first harmonic
(''IPA1'': $a_0 = 1$, ''IPA0'': $a_0 \to 0$).}
\label{fig.5}
\end{figure}     
     
In the model (\ref{eq.convolution}), we employed the independence of the intensity variation 
and energy distribution (frequency bandwidth) for small values of $a_0$ 
and also omitted effects arising from
the shifted threshold $2 m_*$. Although a pair emerges finally with asymptotically
free $e^\pm$ masses $m$, the mass shift during the production of the pair
still leaves imprints on the differential spectra in a pulsed field \cite{Heinzl-PLB}. 
Focussing now on these intensity effects for larger values of $a_0$, one can define 
$\langle \sigma_1 \rangle_{\rm int} = \int d \phi\, g^2(\phi) 
\sigma_1(s, a_0 \to a_0 g(\phi)) / \int d \phi g^2(\phi)$,
where $\sigma_1(s, a_0)$ is the non-perturbative cross section
in IPA related to the first harmonic, i.e.\
$\sigma_1$ from Eq.~(\ref{eq.1}). 
That folding model approximates quite well the spectrum obtained in the 
previous section. For a comparison with numerical results, 
a suitable cut in phase space, e.g.\ $\ell \le 1.2$
(this value is dictated by the lowest gap in the $p_\perp - y$ differential spectrum
for not too large values of $a_0$ and not too short pulses), 
separates here the first FPA harmonic
thus defining $\sigma_1^{FPA}$. In fact,
fig.~\ref{fig.5} clearly exhibits the shift of the threshold from $2 m_*$ towards $2m$
even for $a_0 = 1$, i.e., the IPA threshold $2 m_*$ is replaced by 
the FPA threshold $2 m$ for smooth pulses.
However, already at the IPA threshold $\sqrt{s} = 2 m_*$ the cross section
$\langle \sigma_1 \rangle_{\rm int}$ starts dropping when going to smaller values of
$\sqrt{s}$. This behavior is much more prominent for pulses with
a pronounced flat-top section.  
The shape of the curve $\langle \sigma_1 \rangle_{\rm int}$
as a function of $\sqrt{s}$ again is sensitive to the pulse shape but not to $N$.
The excess at $\sqrt{s} < 2m$ is due to effects of the frequency spectrum
discussed above.

This simple model consideration demonstrates the physics content of the pair production
process in pulsed laser fields by identifying two regimes (i) $N^{-1} > a_0^2$,
where the frequency distribution essentially determines the spectrum, 
and (ii)  $N^{-1} < a_0^2$, $a_0^2 > 1$, 
where the intensity variation in the pulse plays an important role.
Our findings may help to improve simulations of cascade formation
\cite{Bell-2008}
which are based on certain parameterizations of the pair production cross section.

\section{Summary}

In summary we have shown that for short weak and medium-intense laser pulses the generalized
nonlinear Breit-Wheeler process is strongly (order of magnitudes) enhanced
in the subthreshold region $\sqrt{s} < 2 m$. The effect is similar for linear 
and circular polarizations.
While a qualitative explanation was given in \cite{Titov}
for circular polarization, one may attribute the
enhanced subthreshold pair production to the non-monochromaticity of the laser
pulse due to its finite duration. This complements the strong impact of the
temporal pulse shape on the multi-differential spectra found in \cite{Heinzl-PLB}.
Effectively, the high-frequency content of a pulsed laser field leads to a decrease
of the threshold harmonic $n_0$ for given value of $\sqrt{s}$.
The nonlinear (multi-photon) Compton process as crossing channel of the
Breit-Wheeler process has shown an analog sensitivity to the temporal beam
shape \cite{Boca-2009,Mackenroth-2011,NF-96,Seipt-2011}.
Our analysis evidences that pair production in pulsed laser fields depends
sensitively on both the frequency spectrum and the intensity variation in the
course of the pulse. Modelling the pulse by a box shape one would ignore the
intensity variation which is particularly important for stronger laser
fields with $a_0 \gg 1$ and higher harmonics with $n \gg 1$.  
The pair production is an important step in the formation of QED cascades
\cite{Bell-2008} which are thought to become important in high-intense laser fields.    

The authors gratefully acknowledge stimulating discussions with T.Heinzl, A.~Hosaka,
H.~Ruhl, and H.~Takabe. We thank T.~E.~Cowan and R.~Sauerbrey for their supportive 
interest. 

\end{document}